\documentclass[conference]{IEEEtran}

\usepackage{hyperref}
\usepackage{booktabs}
\usepackage{framed}
\usepackage[utf8]{inputenc}

\usepackage{acronym}
\acrodef{DNM}{Darknet Marketplace}
\acrodef{FE}{Finalize Early}

\usepackage{xparse}
\NewDocumentEnvironment{dnmquote}{o}{
  \noindent
  \begin{quotation}\sffamily\itshape
}{
  \end{quotation}
}

\begin{document}

\title{How darknet market users learned to worry more and love PGP: Analysis of security advice on darknet marketplaces}

\author{
\IEEEauthorblockN{%
Andrew C. Dwyer\IEEEauthorrefmark{1},
Joseph Hallett\IEEEauthorrefmark{2},
Claudia Peersman\IEEEauthorrefmark{2},
Matthew Edwards\IEEEauthorrefmark{2},
Brittany I. Davidson\IEEEauthorrefmark{3},
Awais Rashid\IEEEauthorrefmark{2}}
\IEEEauthorblockA{\IEEEauthorrefmark{1}Durham University}
\IEEEauthorblockA{\IEEEauthorrefmark{2}University of Bristol}
\IEEEauthorblockA{\IEEEauthorrefmark{3}University of Bath}
}

\maketitle
\begin{abstract}
  Darknet marketplaces, accessible through, Tor are where users can buy illicit goods, and learn to hide from law enforcement.  We surveyed the advice on these markets and found valid security advice mixed up with paranoid threat models and a reliance on privacy tools dismissed as unusable by the mainstream.
\end{abstract}

\Acp{DNM} are places where illicit goods, such as narcotics, stolen credit card details, and weapons, are bought and sold. In order to facilitate the sale of such goods, \acp{DNM} host forums that serve as a community for users to exchange ideas, report information on other users, markets, and products, and crucially how to protect themselves, both from market participants who are identified as \emph{scammers} as well as law enforcement.

Whereas traditional cybersecurity advice looks to advise organizations~\cite{such2019basic} and individuals~\cite{ncsc2016threerandomwords} on how to protect themselves from criminals online, on \acp{DNM} these roles are reversed.
Mainstream cybersecurity does not typically include hiding data from forensic investigators and law enforcement.
Most users of the mainstream internet---albeit with some significant exclusions---will not face a prison sentence if their activity is uncovered.
This makes the guidance offered on \acp{DNM} of particular interest, partly for law enforcement to better understand the security capabilities of users, but also because \ac{DNM} users are a community who remain relatively distinct from other internet users: they have their own preferred tools, techniques, and threat models.

From a security and privacy perspective, this makes these communities interesting: in \acp{DNM} there is a community of internet users seeking to work outside of the reach of law enforcement and dominant societal norms.
We are interested in their security and privacy practices to better understand how these communities operate, and in order to help law enforcement to intervene to close these markets when they are actively funding cyber-criminal enterprises. 
So what are \ac{DNM} users learning from forums hosted on these sites? Is cybersecurity `advice' correct? And what threat models motivate and inform \ac{DNM} users to offer guidance?  In this article we present an analysis of several \ac{DNM} forums to detail the guidance they are offering and receiving, and the threat models that underlie them.

\section{Source Data and Analysis}

\begin{table*}
  \centering
  \caption{Marketplaces examined in this study (data taken from~\cite{gwern2013darknet}).}
  \sffamily
  \begin{tabular}{l l ll l l  l}
    \toprule
    Name      & Users & Activity Period & Observations           & Closure Reason         & Goods available                                 \\
    \midrule
    Andromeda & 270   & 2014 Apr--Nov   & 270 (2014 Jan--Dec)    & Scam (No arrests)      & Drugs, Guns, Credit cards                       \\
    Agora     & 365   & 2013--2015      & 2,232 (2014 Feb--Mar)  & Voluntary (No arrests) & Drugs,  Guns\footnote{Until 2015}, Credit cards \\
    DHL       & 46    & 2015--2017      & 190 (2015 Apr--June)   & Hacked (No arrests)    & Drugs                                           \\
    Havana    & 36    & 2015 May--June  & 1,212 (2015 Apr--June) & Hacked (No arrests)    & Drugs                                           \\
    Pandora   & 5,168 & 2013--2014      & 50,298 (2013--2014)    & Scam (No arrests)      & Drugs, Guns                                     \\
    \bottomrule                                                                                         \\
  \end{tabular}
  \label{tab:markets}
\end{table*}

\begin{table*}
  \centering
  \sffamily
  \caption{Posts from each of the communities analyzed after each of the filtering steps.}
  \begin{tabular}{l r r r r r}
\toprule
Community & Initial & Filtering & Tech  & Advice \\ 
\midrule
Agora     & 2,232   & 402       & 248   & 41     \\ 
Andromeda & 270     & 140       & 110   & 33     \\ 
DHL       & 190     & 47        & 33    & 15     \\ 
Havana    & 1,212   & 56        & 54    & 30     \\ 
Pandora   & 50,298  & 7,251     & 3,805 & 2,153  \\ 
\addlinespace
Total     & 54,826  & 7,896     & 4,250 & 2,272  \\ 
\bottomrule                                              \\
\end{tabular}

  \label{tab:filtering}
\end{table*}

To explore what guidance was available, we analyzed forum posts from the five \acp{DNM} (Andromeda, Agora, Pandora, DarknetHeroes League (DHL), and Havana) archived by the \emph{DNM Archives}~\cite{dnmArchives} (Table~\ref{tab:markets}). These cover a range of marketplaces from 2013 through to 2015, with variety of \ac{DNM} size and longevity. The data for each of the \acp{DNM} consists of scraped text from the body of the posts, as well as subject lines, author information, and post history.

Though there are 54,826 forum posts from each of the \ac{DNM} forums, not all posts discuss cybersecurity and fewer still offer advice.
To analyse only those which offer advice, it was necessary to reduce the number of overall posts in the dataset.  This was done in 3 steps:  first by \emph{filtering} based on security and privacy related keywords derived inductively from the data, second through a manual analysis of nearly 8,000 posts to identify \emph{technology} focused posts, and third in a review of those posts to identify those offering \emph{advice}.
The distribution of the data from each of the marketplaces after each filtering step is shown in Table~\ref{tab:filtering}.

To analyze the data taken from the \acp{DNM}, we use a qualitative coding-based approach, combined with quotes to illustrate the various themes we identify in the data.  There is surprisingly little work exploring what and how people can learn from \acp{DNM} and how knowledge is transferred in these communities---our analysis shows a community that are actively teaching each other how to use security and privacy tools.  By better understanding the challenges they face and approaches they take we can help make our security and privacy tools more usable for all users---not just the cyber-criminal ones---and help law enforcement better understand how to disrupt these communities when they are being used for illicit trade.

\section{What advice is there on darknet marketplaces?}

So what advice is there on the various \acp{DNM}?
Using a qualitative approach we repeatedly reviewed \ac{DNM} posts to develop a \emph{codebook}: a list of descriptive terms that capture the content and themes within the \ac{DNM} posts.

\begin{table*}
\centering
\caption{Code book, with 24 different categories used on the re-coding of the dark net marketplace forum posts.  Some posts were described using multiple codes.  No more than 12 codes were used to describe any individual post.}
\label{tab:codebook}
\sffamily\small
\begin{tabular}{l r p{\dimexpr 0.7\linewidth-2\tabcolsep}}
\toprule
 {Code} & Count & {Description} \\
\midrule
 \textbf{Transactions} & 1126 & Reference to any mention of the movement of money, the use of \emph{Finalizing Early} (FE), and escrow (sometimes escroll) as well as refunds from the DNM (sometimes \emph{refund center}). \\\addlinespace
 \textbf{User Review} & 1063 & Reference to any individual DNM user or a call for users to conduct research before engaging on a DNM. \\\addlinespace
 \textbf{Vendor Scam} & 871 & When a vendor (who is also a user) does not ship a product, does not process a refund, or is a called a \emph{scammer} by a user. \\\addlinespace
 \textbf{Support} & 729 & Reference to any member of \emph{adminstration} or support of a DNM or methods where support may be required to act (such as over user removal, verification, or particular actions to support a user). \\\addlinespace
 \textbf{Platform Review} & 663 & When a forum posts refers to a DNM's functionality, appearance or bugs and errors (this can also include suggestions for platform improvements). \\\addlinespace
 \textbf{Cryptography} & 619 & Any reference to the use of cryptography. \\\addlinespace
 \textbf{Crypto-Currency} & 483 & When a forum post refers to the use of cryptographic wallets, or when a crypto-currency (mostly Bitcoin) is a focus of the post (but not when it is merely mentioned, such as talking about BTC losses). \\\addlinespace
 \textbf{Compromise} & 423 & Reference to any form of technical \emph{compromise} in terms of either a hacker, exploitation of a vulnerability, or warning of one (such as phishing, clicking on unknown links, and so on). \\\addlinespace
 \textbf{Non-Technical Trust} & 403 & Any reference to trust, honesty, reliability and so on which is not based on a technological form of trust (such as cryptographic verification of identity). \\\addlinespace
 \textbf{Verificiation} & 344 & Any reference to the technologically-mediated confirmation of identity, often used in conjunction with the use of Crypto. \\\addlinespace
 \textbf{Tools} & 320 & Mention of any technological tools, including, but not limited to, operating systems, crypto-currency tumblers, and software and code packages. \\\addlinespace
 \textbf{Platform Scam} & 289 & Reference to a platform or support conducting a scam, compromising accounts, or stealing money from users. \\\addlinespace
 \textbf{Authentication} & 256 & Targeted code around the use of a DNM account, log-in, and passwords. \\\addlinespace
 \textbf{Legal} & 196 & Any reference to interception and seizure (often of shipped products), government bodies, law enforcement and authorities, or a government intervention to a DNM. \\\addlinespace
 \textbf{Anonymity} & 191 & The explicit reference to the concept of anonymity or where anonymity is the implied consequence of a piece of advice (such as to keep information private). \\\addlinespace
 \textbf{Market Availability} & 178 & Any reference to the stability of a market, its closure, or its current online status. \\\addlinespace
 \textbf{Product Review} & 91 & Any explicit review of a product or service. \\\addlinespace
 \textbf{Multi-Sig} & 72 & Any reference to the use of \emph{multiple signature} technology that is often used in conjunction with the protocolls of crypto-currencies. \\\addlinespace
 \textbf{User Scam} & 68 & When a user (who is not a vendor) is called a \emph{scammer} (which can also include the impersonation of another vendor either on the market's forum or from another DNM). \\\addlinespace
 \textbf{Vulnerability} & 31 & Reference to a technological vulnerability that could be exploited (and therefore bugs, unless they fall into this category, belong to Platform Review). \\
 \textbf{Physical Security} & 30 & Typically this referes to \emph{physical} addresses, threat of violence, or prisons. \\\addlinespace
\bottomrule \\
\end{tabular}
\label{tab:final_code_book}
\end{table*}

Table~\ref{tab:final_code_book} shows the distribution of codes used to capture topics from the \ac{DNM} forum posts offering advice.
As is likely to be expected for a marketplace, discussion of \emph{transactions}, possible \emph{vendor scams}, \emph{reviews} of users and markets, and administration \emph{support} requests are common.
Also common is discussion of how marketplace users can protect themselves, not just with \emph{cryptography} and \emph{privacy tools}, but how to operate safely in an environment where vendors, users and the markets themselves are routinely untrustworthy and law enforcement is an active adversary.
For example, one user responds in a thread started by a new user titled \textit{``If i do deposit BTC to my pandora account..... Will it show up................?''} advises:
\begin{dnmquote}[1841]
  ``People need to realise that you need to be security conscience Every.Single.Time. you use these sites!  EVERY SINGLE TIME. Transfer a couple of dollars to test response times, passwords etc and then transfer them out again. Once you feel comfortable/confident enough then send your big dollars over, order immediately and then transfer out any change, leave a zero balance. EVERY SINGLE TIME.    It hurts getting burnt, it hurts less if it only happens once.''
\end{dnmquote}

\subsection{Security and Privacy Tools}
Others advise users how to use tools.
Posts describing the use of Tor (usually through the Tails Linux distribution), VPNs, and Bulletproof hosting are common; but of particular interest is the use of PGP.
The use of PGP encryption is routine on \acp{DNM}, unlike on the open internet. Despite the well known usability challenges of PGP~\cite{whitten1999johnny},
users and vendors advise each other which clients to use:
\begin{dnmquote}[125]
  ``Ask the magical unicorn that killed my milk man...  Seriously though, if you want a pgp, you need to download a program that will make store encrypt and decrypt your PGP stuff. If youre using windows, look up Kleopatra..''
\end{dnmquote}
We also identified longer tutorials on how to set up PGP running off USB sticks.  Sometimes these posts would come with offers of support, for example:
\begin{dnmquote}[130]
  ``Step 1: buy a \$5 USB stick Step 2: download gpg4usb onto the stick
  [\ldots]
  Step 11: put your USB stick in a 'safe' place Step 12: get high...  ...Very easy - PM me and i'll give you more detail if you can't follow it...''
\end{dnmquote}
\begin{dnmquote}[3842]
  ``HI Guys let me assist you here,  After becoming fed up of teaching PGP to customers and vendors alike I decided to write a simple guide to walk you through the process of PGP encryption. I posted it on the forums this morning to help those users that have difficultties using the application. I saved a copy to hand out to any of our customers that need some PGP guidance. Here is my idiot  proof guide, just follow the 20 simple steps to achieve success
  [\ldots]''
\end{dnmquote}
The need for these tutorials, however is very much driven by users: many struggle to understand how PGP works (as might be expected):
\begin{dnmquote}[3418]
  ``Hey everyone basically im after some help to do with my pgp key I just cantvget my head round it and more I read bout it more it fucks with my head is there any one who could help me or shed bit of light on this topic fir me much appreciated eddy!!;-)''
\end{dnmquote}
In response, the user was given a tutorial:
\begin{dnmquote}
  ``[PGP Tutorial!!!]  I'll walk you through the process of getting started with a very common and easy to use program. [\ldots]''
\end{dnmquote}
Yet despite the clear signs 
that show users consistently struggle with PGP (and sheer number of \emph{PGP is hard} research papers~\cite{sheng2006johnny,whitten1999johnny,ruoti2016we}), many users marketplace vendors and administrators insist that PGP is easy to use:
\begin{dnmquote}[588]
  ``We encourage everyone to uses PGP for everything. The only thing that truly protects your sensitive information is encryption. On any market there can be no way of being safe without ensuring you are fully protected and PGP is your best way to do that.  Using PGP is easy and it is multi-platform. anyone capable of using bitcoin is capable of using PGP.   PLEASE LOOK IT UP AND LEARN IT. we cant stress this enough.
  [\ldots]
  STAY SAFE!''
\end{dnmquote}
\begin{dnmquote}[589]
``hey  PGP/GPG isn't too difficult once you get your head around what is happening.
[\ldots]''
\end{dnmquote}

\subsection{Trust}

For what then are the \ac{DNM} users using PGP?
Whilst PGP is regularly used for encrypting messages between vendors and users, it is also used for other purposes---including as account authentication for the forums themselves. One notable use, however, is for verifying the identity of vendors between markets.  Because \acp{DNM} can come and go relatively quickly, a high profile vendor on one market may also wish to sell on other marketplaces: sometimes scammers exploit this by copying or imitating usernames of other vendors.  To defend against this threat, users attempt protocols where they challenge suspect vendors to decrypt messages with older keys or keys from other markets.  If a vendor cannot do so then users encourage other users to consider them \emph{unacceptably} untrustworthy.
\begin{dnmquote}[1902]
  ``[\ldots] Please remember: Some scammers are using reputable dealers from SR and BMR. (always verify original PGP keys to make sure the vendoe is not an imposter) [\ldots]''
\end{dnmquote}
\begin{dnmquote}[2083]
  ``[\ldots]
  All C buyers STOP buying C until you verify the sellers.  Use the original PGP public keys posted on BMR.  you will find that many of the vendors with same usernames are imposters.  Do Not Buy anything until you verify using original keys.  Do NOt accept any excuses.  Real legit dealers do not lose their original keys or change them.''
\end{dnmquote}
To mitigate this threat some marketplaces claim to \emph{verify} vendors, yet this verification process isn't entirely trusted:
\begin{dnmquote}[1064]
  ``Just pay no attention to the "verified vendor" label, and just because you see their public pgp key on their profile, that doesn't mean anything, anyone can copy and paste that.  Sure they won't be able to decrypt your dispatch address, but then if they have no intention of dispatching anything, it matters not.
  [\ldots]
  Be 110\% certain it is the real vendor not a fake. Send some encrypted messages and grab their PGP key from another marketplace forum like the green camel.''
\end{dnmquote}

\ac{DNM} users use PGP as one mechanism to decide whether to \emph{socially} trust new vendors through establishing forms of technical verification, however there is also constant suspicion and questioning as to whether any vendor is trustworthy, and the legitimacy of prospective vendors is actively discussed.  For example, one user sees a new vendor and queries whether they are \emph{too good to be true}.
\begin{dnmquote}[271]
  ``I have looked and searched some more and finally found a USA vender for opium his/her vender name is Painfree and says he/she has been on other markets. Is this to good to be true idunno. Only like one feed back. Answers would be gratefully appreciated!!''
\end{dnmquote}
Another user comments, highlighting points of suspicion:
\begin{dnmquote}
``Painfree looks really suspicious. Those prices are pretty suspicious, and I wouldn't trust a vendor without a profile description or a PGP key. If you at his listings, he claims to have lost too much on SR [Silk Road] and BMR [Black Market Reloaded, not analyzed in this study], so he doesn't trust escrow, but he seems to have a confusing escrow policy. But I don't remember him at all, and he isn't verified or anything. Plus he has a 'limited time sale' on all his products. And the picture he has of the opium looks awfully suspicious as well.  [\ldots]''
\end{dnmquote}

Here the lack of PGP is used as a sign of bad faith.  The vendor does not have one, so they may not be reputable.  Vendors (and to a lesser extent users) are expected to demonstrate technical competence when trading. Along with other signs, including opting out of escrow mechanisms (purportedly) designed to protect customers, the vendor is marked out as being suspicious and a potential scammer.  This example suggests that PGP is not always perceived as a purely technical mechanism for validating identities and encrypting messages: sometimes the act of showing that you know the tools to stay secure builds the customer's confidence---even if they may not use them.

\section{But is the advice correct?}

Given the focus on privacy and security advice on the various \ac{DNM} posts, one might reasonably ask whether the guidance is any good---that is, does the advice actually help anyone stay secure or protect their privacy?
For the most part, the advice we identified on the \acp{DNM} was `correct'---by which we mean it described an appropriate way to mitigate a threat---however sometimes the extent of the recommendation in order to protect a user's identity could be considered extreme. For example one post (477) recommended in order:

\begin{itemize}
  \item Disabling JavaScript in case of zero-day attacks on the browser.
  \item Avoiding the use of any darknet site which uses JavaScript.
  \item Avoiding clearnet sites while using Tor.
  \item Avoiding clearnet sites which use Javascript.
  \item Disabling browser support for \texttt{iframes}.
  \item Using strong passwords.
  \item Being wary of attackers brute forcing CAPTCHAs.
  \item Storing passwords only in a PGP encrypted textfile.
  \item Avoiding accessing sites at predictable times to avoid leaking location data.
  \item Using a VPN to help obfuscate when you connect to sites.
  \item Varying writing style to prevent stylometry attacks.
  \item Avoiding writing on the clearnet at all (especially Facebook and Twitter) to reduce attack surface for stylometry attacks.
  \item Not searching for information on the darknet on the clearnet.
  \item Not using Google on the clearnet.
  \item Deleting your past writing from the clearnet.
  \item Deleting all social media and all information about you on the clearnet.
\end{itemize}

For the most part, this advice was targeted towards users within a \ac{DNM} that focused on buying drugs (Andromeda).  Much of the advice is reasonable---we advise even non-darknet users to store passwords securely; separating day-to-day use of the internet from illegal activity is sensible---and yet together the advice could be understood as somewhat extreme.  Disabling Javascript fully on the clearnet (unfortunately) leads to sites breaking.  Whilst stylometry attacks exist, instead of recommending to not post on darknet forums, users are encouraged to remove writing from the clearnet and not using social media. This is odd as this might make a \ac{DNM} user rather more notable against the majority that do.  A simpler solution might be to just not use or post on \ac{DNM} forums; yet that is not what they recommend.  Instead it is the user's interactions with the \ac{DNM} that must be protected. \emph{If} that comes at the cost of more normal social interactions; then that is reasonable.

Is any of this advice \emph{wrong}, though?  Security is not an absolute property and we did not attempt to judge whether advice was appropriate but rather identified  examples where it was \emph{technically incorrect} (that is, where it would not have the desired effect).
Whilst there were very few \emph{absolutely incorrect} pieces of advice, there were several themes among the advice that were more questionable:

\begin{itemize}
  \item Assertions that one operating system (usually Linux) was inherently more secure than any other.
  \item Assertions that it was safe to \ac{FE}---many markets offer an escrow service where the market will arbitrate in the case of disagreements between vendors and buyers (such as the non-receipt of drugs).
    \ac{FE} allows buyers to opt out of this arbitration process (and vendors to be paid quicker), but also allows vendors to steal the buyer's money and run.
    As one buyer noted:
    \begin{dnmquote}[4320]
      ``Having had maybe 20/30 sucsefull transactions total on darkweb. In the last month i've been skanked twice (both times on Pandora (Stealth100, classatrader), and both vendors had seemed ok at the time, so i FE'd.  Now I am going to refuse to do FE. [\ldots]''
    \end{dnmquote}
    However, it may also be questionable to trust the escrow service provided by \acp{DNM}, due to the potential of administrators being able to close down the marketplace and `scam' its users by stealing that held in escrow. This demonstrates the intricate balance of risk and trust that are present for all users of \acp{DNM}. 
\end{itemize}

Whilst different operating systems do have different security features, to assert one is more secure than another uncritically is incorrect, as it is more often the ecology of its \emph{use} that affects its security, rather than any inherent property.  An exception could be made, however, for ephemral OSs---such as TAILS---though there is little empirical evidence to support this, and even these OSs can be used insecurely.  Assertions that it is safe for buyers to \ac{FE} are false: escrow is a consumer protection mechanism built into some markets designed to offer some protection to a user at the expense of the vendor, and so opting out of it is of no benefit to buyers and increases the likelihood of a vendor turning scammer. Yet, escrow also transfers trust from the vendor to the administrator.
An administrator in collaboration with a vendor could still scam a user, but these escrow mechanisms shows the complexities of security when there is widespread mistrust.

\section{So what are they so concerned about?}

With \ac{DNM} users discussing what could be considered an extreme (compared to non-darknet users) approach to security and privacy; what are the \ac{DNM} users' threat models?  Who do they think is more likely to cause them some form of harm and their capabilities?
The long list of advice in the previous section also comes with a warning:
\begin{dnmquote}[477]
  ``[\ldots]You should expect that any observing persons have very complicated algorithms for searching the clearnet and correlating posts made there with posts you made here. They have access to Google and all the other search engines, so additionally do not search for anything incriminating.[\ldots]''
\end{dnmquote}
The poster believes that there is an advanced attacker who is actively seeking them, has access to large internet company records, and is attempting to work out who they are personally based on stylometry and their web history.

Another poster considers what would happen if their messages (and purchase history) were leaked. They consider how their posts and messages might be used to incriminate them in a court of law.  Their mitigation to this is to use PGP everywhere to encrypt their messages.
\begin{dnmquote}[1957]
  ``[\ldots]
  Guess what guys, in the event that LORD JEBUS forbid, something goes wrong for you personally any data in plain text will be used by prosecution and/or law enforcement to do one or more of the following, among other things:   1)  (Further) evidence your involvement in whatever by way of often detailed affirmative in-plain-text written confirmations of whatever.  They will be read in court.   2)  They will be read in court sometimes for the sole purpose of building a character profile of you to reference to a jury.   3)  Semantically identify you which may lead to your being found IRL and in other places on the Internet.
  [\ldots]
  Once you consider other parties accessing communications beyond just law enforcemet (think administrators, hackers, journalists, phishers, whoeverthafucks), PGP use in all cases is ever and evermore justified. THERE ARE NO REASONS ONE OUGHT NOT USE PGP IN ALL CASES.
  [\ldots]''
\end{dnmquote}

One user described how they access \acp{DNM} at work. They articulate how they have considered a threat (their employer monitoring them using an internal hardware-based keylogger), and how they decided that the threat is relatively low.
\begin{dnmquote}[4539]
  ``Dude, install TAILS to a USB. It's way better. I am even confident now to order via my work pc (obv not on thier network ;-) ) - the only way they could find out how naughty I am is with a hardware keylogger installed inside the laptop - and i'm not that paranoid.  [\ldots]''
\end{dnmquote}

Another user wrote a 2,000 word guide for drug vendors on how they should assess threats and consider the threat of law enforcement, saying:
\begin{dnmquote}[2318]
``[\ldots]You need to take the time to not just think about what your doing, but
to visualize it. Visualize each step, and try to imagine all possibilities with
each step, such as how the police may evidence or identify you with that
particular step in your shipping/receiving process. Then make the appropriate
changes.[\ldots]

[\ldots]
foil LE's ability to follow
a mailing back to it's origin, and prohibit LE from compiling a list of buyers
from your mailing labels [\ldots]''
\end{dnmquote}
They go on to recommend a number of tips for vendors, from buying new wireless adaptors weekly (to avoid MAC addresses being associated with them) to a guide on how customs agents x-ray packages, including:
\begin{dnmquote}[2318]
``[\ldots]Sellers can just toss a little hair from the local Hair Solon in the pack, send
LE on a wild goose chase LOLOL

Every measure you can take to obscure your identity and location should be done.[\ldots]''
\end{dnmquote}
Whilst the advice is idiosyncratic, the process described to consider all
possible threats in each step of a vendor's selling process, to take
appropriate steps to ensure they obscure their identity, and to change their
processes is sound.  Most of the techniques they describe (including adding
hair) obfuscate their identity rather than provide security, but their threat
model requires them to avoid law enforcement ever suspecting that they may be dealing drugs,
rather than giving deniability should law enforcement investigate.

\section{Threat Models and Tor}

\begin{figure*}
  \centering
  \begin{framed}
    \begin{itemize}
    \item\textbf{Client side scripting.} Code inside or outside the browser may be used to deanonymize.
    \item\textbf{Browser fingerprinting.} Unique features of a particular setup may deanonymize users.
    \item\textbf{Side-channel leaks.} Input provided by users (e.g. login details, payment information) may lead to deanonymization or credential theft.
    \item\textbf{Node operation.} Actors may operate Tor nodes maliciously.
    \end{itemize}
  \end{framed}
  \caption{Summary of threat model for Tor identified by Gallagher~et~al{.}~\cite{gallagher2017new}.}
  \label{fig:gallagher}
\end{figure*}

Given that a \ac{DNM} user must typically also be a Tor user to access the market itself, it seems reasonable to compare the two groups.  Not all Tor users are also \ac{DNM} users, but there has been some work studying general Tor user behavior.
Gallagher~et~al{.} studied the extent expert and non-expert users understood the threat model behind Tor~\cite{gallagher2017new}.  They found that whilst experts understood the threat model they identified (Figure~\ref{fig:gallagher}), non-experts conflated it with other threats.
To what extent do \ac{DNM} users seem to assimilate Gallagher~et~al{.}'s general Tor threats, and where do they go beyond these and add threats specific to their use of \acp{DNM}?

Much of the guidance on darknet markets sought to address the threats in Gallagher~et~al{.}'s Tor threat model---with guidance to disable Javascript, use standard TAILS environments, and to enforce a clear separation between \ac{DNM} user's darknet and mainstream internet identities.

But the advice identified in this paper for darknets also goes beyond Tor to include dealing with untrustworthy vendors and markets, as well as law enforcement.  In addition to the \emph{technical} threats that Gallagher~et~al{.} describe, \ac{DNM} users must also address legal and social threats which has been seen in cases of civil dissent elsewhere~\cite{akbari2019platform}.

\begin{figure*}[t]
  \centering
  \begin{framed}
    \begin{itemize}
    \item\textbf{Don't trust the vendors or the markets.}  Vendors do not have customers' best interests at heart, and it is more a question of when they defraud the customer, not if. Customers mitigate by using marketplace escrow features, never finalizing early, and never putting anything into the market that they cannot afford to lose.  Assume any links are to backdoored software or phishing sites.
    \item\textbf{Encryption and obfuscation.}  Messages between customers and vendors are of interest to law enforcement as they can contain data (including stylometric data) about vendor and customer.  The simplest mechanisms to protect against the interception of this data is to use PGP for all communication and to verify keys.  Note that this does not protect against one side releasing their key to law enforcement.  
      \item\textbf{Plausible Deniability.}  If law enforcement have evidence that illicit activity has taken place, then they may start to build a case that leads to an arrest.  Mitigate with RAM-based ephemeral OSs such as TAILS and memorizing URLs to marketplaces instead of writing them down.
    \end{itemize}
  \end{framed}
  \caption{Additional threats for \ac{DNM} users.}
  \label{fig:gallagher++}
\end{figure*}

Figure~\ref{fig:gallagher++} describes three additional threats that \ac{DNM} users include in their threat model.  These include limiting trust in vendors and the market, as well as adding layers of obfuscation and deniability.  Whilst some of these threats are mitigated with purely technical approaches, like Gallagher~et~al{.}'s: for example using TAILS, they also start to rely on approaches which require them to modify their behavior: for example memorizing URLs instead of writing them down, lest law enforcement link their knowledge of drug forums to their use of them.

Prior work on the use of Tor has either focused on the technical design behind the Tor network itself~\cite{dingledine2004tor,sanchez2017onions} or on people seeking anonymity for privacy reasons~\cite{gallagher2017new,forte2017privacy} (as opposed to people seeking anonymity to commit criminal acts).  Our exploration of \acp{DNM} suggests that, especially for \ac{DNM} users, the use of Tor is part of a larger set of goals, and Tor is just one part of a larger threat model.  There is an entire ecosystem of privacy-focused users developing guides and tutorials for their own community that appears to be largely distinct from mainstream privacy: using much of the same software, yet with different behaviors and threat models. These users' privacy practices deserve further study.

\section{Connections between privacy communities and darknet users}

In the data from the \ac{DNM} forums, several examples of users offering advice of how to stay \emph{safe} using Tor and PGP were identified.  Yet these tools are also used by the wider privacy and cryptography communities: why are they rewriting tutorials explaining how to use PGP and Tor when there are already many guides written by these other communities?  The use of software from these communities is routine, yet we saw few links to their guides with forum members advising that they create their own.

\begin{dnmquote}[640]
  ``If we add here in the Absolem Forums a Tutorial part we can share our know-how to help around people would like to learn.  Maybe a tutorial part in the forum for security tutorials (VPN,SOCKS and JS) for anonymity. A tutorial how to use Bitsigner and the TorBrowser etc. \ldots''
\end{dnmquote}

Many of these requested tutorials also exist on the regular internet.  For example organizations like the Electronic Frontier Foundation offer tutorials on PGP and using Tor safely\footnote{Available: \url{https://ssd.eff.org}}---yet external references are not included on \ac{DNM} forums.  We can speculate as to why these tutorials are not referenced:  perhaps \ac{DNM} users see this as being part of \emph{plausible deniability} (Figure~\ref{fig:gallagher++})---if they have not seen a Tor tutorial then how are they using the darknet marketplaces?  An alternative explanation is that these tutorials represent a \emph{club good}~\cite{sandt2019deviant,buchanan1965economic}: a mechanism whereby the users of the \ac{DNM} forums produce peer-reviewed guides within their own \ac{DNM} group.  Whilst membership of \ac{DNM} marketplace forums is often open; the requirement to find the forum and access it through Tor effectively excludes some potential members. The darknet privacy tutorials are written for darknet marketplace members to help make the marketplace more secure (and infiltration by law enforcement less likely). Why reference external guides, and use risky hyperlinks, when it is possible to keep it within the club?

The counterpoint to this, however, is that whilst \ac{DNM} users produce their own guides, they do not do the same for software.  The advice on \ac{DNM} forums recommends using relatively standard, if unpolished, open source privacy software such as \emph{GPG4Win} and \emph{Kleopatra} for PGP, and the \emph{TAILS} live CDs or \emph{Tor Browser Bundle} for accessing the darkweb.  One explanation could be that there is a larger barrier to entry in writing software than there is in writing a tutorial, especially cryptography software where the general advice even for experts is \emph{``Don't roll your own crypto''}.  Yet darknet markets do have technical expertise; there are users who set up VPNs and proxies for users, as well as mail servers explicitly for use with \acp{DNM}---and these services are recommended, so long as the other software is layered on top of them.

Perhaps the reason for not recommending their own software, but conversely using their own guides, relates to  trust.  \acp{DNM} are routinely places where scams are conducted, administrators are untrustworthy, and are taken down by law enforcement.  Some of the most frequent categories in our codebook of the different kinds of advice on \ac{DNM} (Table~\ref{tab:final_code_book}) relate to users advising each other to do research before engaging on a darknet marketplace (the \emph{user review} code) or reporting scams (the \emph{vendor scam} category).  Given the high likelihood of a scam on the marketplace, perhaps trusting software to be non-malicious is too much.  Steps in a guide can be compared to other guides to vet for trustworthiness and a user can always opt to deviate from the advice; but the same is not always possible for software, even open source code.  Users may then see the benefit of using something from outside their own ecosystem that is understood to be secure with defined and tested parameters, where the chance of a scam is lower and is easily integrated within their security and privacy practices. However, the combination of software and practices is unique to this community, which necessitates specialised guides in order to navigate and stay secure on \acp{DNM}

\section{Implications for law enforcement}

Given the illicit nature of \acp{DNM}, what are the implications for law enforcement seeking to disrupt these markets?   First, although the cybersecurity advice on marketplaces was largely valid given an appropriate threat model; advice on how to use tools to access markets was required because users found using various technologies (Tor, PGP, and bitcoin) hard.

\begin{dnmquote}[987]
  ``I found the PGP system in Tails fairly easy. I can't believe how much time I wasted putting off using the Darknet markets because I was put off by the technical side of getting BTC, setting up encrypted wallet on a live USB Tails USB, PGP conversation etc.  However in the end it just took a week or so of reading and setting up in the evenings after work...like barely an hour or so each evening.  I used a PGP on Tails tutorial which I'd saved from the SRv1 forums.''
\end{dnmquote}

Whilst there is a barrier to entry, the tutorials produced by other \ac{DNM} users help and are read by novice users.  Disrupting these tutorials may make it harder for a user to go from being interested in \acp{DNM}, to actively participating.

Deterring new users from becoming active participants on \acp{DNM} is one approach, but how could law enforcement disrupt the vendors on these marketplaces?  Because the advice offered on the market is, broadly speaking, correct or at least appropriate given the users' threat models, disrupting the market through technological means may not be the most effective---though there may be some \emph{low hanging fruit} given the number of vendors and buyers appearing to struggle with the technology.  An alternative approach may be to exploit the \emph{lack of social trust} within the markets themselves.  Our codebook (Figure~\ref{tab:codebook} highlighted a large number of posts where users advised each other to conduct research, or reported vendors or \acp{DNM} as scam marketplaces.  Working to increase the senses of unease in trust between participants on a marketplace may help to move the users and vendors along, and force the vendors to set themselves up again on new markets.  Vendors use public key signatures to assure their customers and transfer their reputations between markets: campaigns about how vendors are working with law enforcement and negative reviews may help push a vendor into the \emph{scammer} category on an already \emph{untrustworthy} market, disrupting their ability to sell and damaging business opportunities.

\section{Conclusions}

\ac{DNM} users are learning about privacy tools and techniques from tutorials written on darknet forums.  Whilst the tutorials are, given a \ac{DNM} user's threat model, correct; the threat models that these users inhabit go beyond a typical Tor user's threat model, including social and trust relationships designed to impede law enforcement.  Perhaps most remarkably of all, darknet marketplaces represent a place where PGP---considered too \emph{unusable} on the mainstream internet---is used regularly and actively recommended.

Ultimately relationships on the darknet between buyers, vendors and the markets are interlaced through relationships to trust.  PGP is the tool they use to technically secure themselves and establish trust relationships on these markets, but the guidance we saw went beyond purely technical relationships---into social relationships, guidance and discussions given between users.  Future work should explore these \emph{social} relationships further and how they evolve as the \ac{DNM} ecosystem changes over time due to the closure of existing \acp{DNM} and the launch of new ones.

\bibliographystyle{plain}
\bibliography{main}
\end{document}